\documentclass[multi]{cambridge6A}
\usepackage[numbers]{natbib}
\usepackage{rotating}
\usepackage{floatpag}
\rotfloatpagestyle{empty}
\usepackage{amsthm,amsmath}
\usepackage{graphicx}

\renewcommand{\vec}[1]{\mathbf{#1}}
\begin{document}

\alphafootnotes
\author{%
        J. Keeling\footnotemark[1],%
        L. M. Sieberer\footnotemark[2],
        E. Altman\footnotemark[3],%
        L. Chen\footnotemark[4],%
        S. Diehl\footnotemark[2]\footnotemark[5],%
        J. Toner\footnotemark[6]}
\chapter{Superfluidity and Phase Correlations of Driven 
  Dissipative  Condensates}
\footnotetext[1]{SUPA, School of Physics and Astronomy, University of St
  Andrews, St Andrews KY16 9SS UK}
\footnotetext[2]{Institute for Theoretical Physics, University of Innsbruck, A-6020 Innsbruck, Austria}
\footnotetext[3]{Department of Condensed Matter Physics, Weizmann Institute of Science, Rehovot 76100, Israel}
\footnotetext[4]{College of Science, China University of Mining and Technology, Xuzhou, Jiangsu 221116, People’s Republic of China}
\footnotetext[5]{Institute of Theoretical Physics, TU Dresden, D-01062 Dresden, Germany}
\footnotetext[6]{Department of Physics and Institute of Theoretical Science, University of Oregon, Eugene, Oregon 97403, USA}

\begin{abstract}
  We review recent results on the  coherence and superfluidity of
  driven dissipative condensates, i.e., systems of weakly-interacting
  non-conserved Bosons, such as polariton condensates.  The presence
  of driving and dissipation has dramatically different effects
  depending on dimensionality and anisotropy.  In three dimensions,
  equilibrium behaviour is recovered at large scales for static
  correlations, while the dynamical behaviour is altered by the
  microscopic driving.  In two dimensions, for an isotropic system,
  drive and dissipation destroy the algebraic order that would
  otherwise exist, however a sufficiently anisotropic system can still
  show algebraic phase correlations.  We discuss the consequences of
  this behaviour for recent experiments measuring phase coherence, and
  outline potential measurements that might directly probe
  superfluidity.
\end{abstract}

\section{Introduction}
\label{keeling_sec:introduction}
This chapter is dedicated to superfluidity, and its relation to
Bose-Einstein condensation, a topic with a long history.  Many reviews
of the concepts of condensation and superfluidity in thermal
equilibrium can be found, see for
example~Refs.~\cite{huang,Leggett1999a,Hohenberg2000,Leggett:QL}.  The
focus of this chapter is on how these concepts apply (or fail to
apply) to driven dissipative condensates --- systems of bosons with a
finite lifetime, in which loss is balanced by continuous pumping.  We
focus entirely on the steady state of such systems, neglecting
transient,  time dependent behaviour.

Experimentally, the most studied example of a driven dissipative
condensate has been microcavity polaritons (see
Ref.~\cite{Carusotto2013a} and chapters... of this book). However
similar issues can arise in many other systems, most obviously photon
condensates~\cite{Klaers2010}, magnon
condensates~\cite{Demokritov2006a} and potentially exciton condensates
(although typical exciton lifetimes are much longer than for
polaritons).  Even experiments on cold atoms could be driven into a
regime in which such physics occurs, when considering continuous loading
of atoms balancing three-body losses~\cite{Falkenau2011} or atom laser
setups \cite{Mew97,Rob08,Rob13}.

Experiments on polaritons are two dimensional, and in two dimensions
it is particularly important to clearly distinguish three concepts
often erroneously treated as equivalent: superfluidity, condensation,
and phase coherence. This is because no true Bose-Einstein condensate
exists in a homogeneous two-dimensional system.  Before addressing
superfluidity and phase coherence in the steady state of a driven
dissipative system, we review in
section~\ref{keeling_sec:superfl-phase-coher} the essential ideas of
condensation, superfluidity and phase coherence for systems in thermal
equilibrium.  In section~\ref{keeling_sec:driv-diss-syst} we set up a
generic microscopic model for weakly interacting driven dissipative
Bose gases, and make precise the sense in which these systems are
non-equilibrium.  Section~\ref{keeling_sec:phase-coher-driv} reviews
the connection of these driven dissipative systems to the
Kardar-Parisi-Zhang equation, and explains the absence, for isotropic
systems, of algebraic order at large distances based on this mapping.
We also show that algebraic order is possible in the strongly
anisotropic case. We frame this discussion in the context of current
experiments, which have all been done in the weakly anisotropic
regime. In section~\ref{keeling_sec:superfl-resp-funct} we discuss the
meaning of superfluidity in a driven, number non-conserving setup and
discuss experimental probes. Section~\ref{keeling_sec:vort-driv-diss}
gives a brief account of vortices in such open systems. Conclusions
and challenges for future research are given in
section~\ref{keeling_sec:future-direct-open}.

\section{Bose-Einstein condensation and superfluidity}
\label{keeling_sec:superfl-phase-coher}

Bose-Einstein condensation for a gas of weakly interacting Bosons is a
phase transition associated with the appearance of off-diagonal long
range order (ODLRO)~\cite{yang62}.  This means that correlation
functions such as $\langle \psi^\dagger(\vec r) \psi^{}(\vec
r^\prime)\rangle$ remain non-zero even between distant points, $|\vec
r - \vec r^\prime| \to \infty$.  These correlations indicate the
spontaneous symmetry breaking of the $U(1)$ phase of the condensate
wavefunction, i.e.  writing $\psi = \sqrt{{\rho}} e^{i \theta}$, ODLRO
corresponds to the phase $\theta$ being correlated at distant points.
In such a symmetry broken phase there is a ``phase stiffness'', i.e.
there is an energy cost, $E[\theta(\vec{r})] = (K_s/2) \int d^2\vec{r}
(\nabla \theta)^2$ for phase twists of the condensate. 

A gedanken experiment to determine this phase stiffness is to measure
the change of energy as one imposes a phase twist between the
boundaries of the system.  Alternatively, since phase gradients
correspond to currents, a more practical way to measure the phase
stiffness is to apply a force that tries to drive a current, and
observe the condensate's response.  The behaviour of a condensate in a
rotating container illustrates the role of the phase stiffness very
clearly~\cite{Leggett1999a,pitaevskii03}. The condensate cannot be
made to rotate except by creating quantised vortices, and these cost
energy due to the phase stiffness. Thus, for slow rotation, the
condensate fails to rotate.  Similar behaviour can be seen in a ring
trap, where the core of a vortex can be located outside the
condensate. This is the Hess-Fairbank~\cite{hess67} effect, and is a
defining property of a superfluid. i.e., when a condensate has
non-zero phase stiffness, it becomes superfluid, as seen by its
reduced response to rotation.

However, superfluidity is not equivalent to Bose-Einstein
condensation; as discussed below, in two dimensions, Bose-Einstein
condensation and ODLRO do not exist, yet superfluidity persists.  It
is therefore useful to be able to define superfluidity directly
without reference to condensation.  This can be done, by defining a
superfluid density as the part of the system that fails to respond to
slow rotations.  This definition also clarifies another important
point: except at zero temperature, a system will have both superfluid
and normal components, as thermally excited quasiparticles can respond
normally to rotations.  To identify the superfluid density we must
consider the response function $\chi_{ij}(\vec{q},\omega)$, which
relates the current $\langle \hat{j}_i(\vec{q},\omega) \rangle$ to the
force that induces it, $f_j(\vec{q},\omega)$:
\begin{equation}
  \label{keeling_eq:1}
  \langle \hat{j}_i(\vec{q},\omega)
  \rangle = \chi_{ij}(\vec{q},\omega) f_j(\vec{q},\omega),  
\end{equation} 
where $i,j$ refer to Cartesian components.  The operator $\hat j_i$
appearing here is the standard particle current written in momentum
space;
\begin{displaymath}
  \hat j_i(\vec{q}) = \sum_\vec{k} \hat \psi^\dagger_{\vec{k+q}} \gamma_i(2\vec{k}+\vec{q})
  \hat \psi_{\vec{k}}, \qquad
  \gamma_i(\vec{K}) = \frac{K_i}{2m}.
\end{displaymath}
We consider systems in which this current is conserved, i.e.
$ \partial_t \rho + \nabla \cdot \vec j = 0$, where $\rho$ is the
particle density.  According to Noether's theorem, current
conservation corresponds to the existence of a $U(1)$ phase symmetry
in the Hamiltonian --- this is the same symmetry which is
spontaneously broken on Bose-Einstein condensation.  Considering
static (i.e.  $\omega=0$) long-wavelength (i.e. $\vec q \to 0$)
currents,  the most general response function possible for an isotropic
system is:
\begin{equation}
  \label{keeling_eq:2}
  \chi_{ij}(\vec{q}\to 0, \omega=0) = \chi_L \frac{q_i q_j}{q^2}
  + \chi_T \left( \delta_{ij} - \frac{q_i q_j}{q^2} \right).
\end{equation}
These terms $\chi_L, \chi_T$ describe the response to longitudinal and
transverse forces, i.e. $\vec{f} \parallel \vec{q}$, which occurs for
a potential force, and $\vec{f} \perp \vec{q}$, which occurs for
rotational or magnetic forces.  Current conservation can be shown to
mean that $(q_i q_j/q^2)m\chi_{ij}(\vec q,\omega) = \rho(\vec
q,\omega)$, where $\rho(\vec q,\omega)$ is the single particle Green's
function, so that indeed $\chi_L$ is related to the total particle
number.  The transverse part of $\chi_{ij}$ describes the response to
rotations; therefore superfluidity corresponds to a reduction of
$\chi_T$.  In a non-superfluid, current is parallel to force, which
means $\chi_T=\chi_L$.  The superfluid fraction of a system is
therefore given by $(\chi_L-\chi_T)/\chi_L$, and the normal fraction
by $\chi_T/\chi_L$.

The above definition of superfluidity depends on the
\emph{equilibrium} effect that a system in true thermal equilibrium
has a reduced response to rotational forces.  This effect is
conceptually distinct from the \emph{metastable} persistent flow that
can also be seen in a non-simply connected (e.g. annular)
geometry~\cite{Leggett1999a}. Metastable persistent flow occurs if one
first sets the fluid in motion by rotating the container while above
the critical temperature, and then cools the fluid until it becomes
superfluid. If the container then stops rotating the fluid remains in
motion, as the lifetime for the current to decay is exponentially
large.

It is also important to note that the superfluid density defined above
is a static property of the system; if excited dynamically, it is
always possible to create excitations out of the condensate.  Because
the elementary spectrum of an interacting Bose gas has a linearly
dispersing Bogoliubov sound mode, this sound velocity $c_s$ acts as a
critical velocity~\cite{Leggett1999a,pitaevskii03,Leggett:QL}: for a
defect moving at a lower velocity, no excitations can be created  beyond
the condensed component.  However, at non-zero temperatures,
thermally excited quasi-particles can respond to flow at any velocity,
and a normal component will occur, as discussed above.  In equilibrium
there is a fundamental connection between the existence of a
non-zero sound velocity and the presence of a non-zero superfluid
fraction: as discussed by~\cite{griffin94}, the finite frequency
response function $\chi_{ij}(\vec{q},\omega)$ has the same poles as
the single particle Green's function, and so the finite frequency
generalisation of the superfluid part of Eq.~(\ref{keeling_eq:2}) is
$\chi^{SF}_{ij}(\vec{q},\omega)=c_s^2 q_i q_j / (c_s^2 q^2 -
\omega^2)$.  The fact that this is finite at $\omega=0, \vec{q} \to 0$
is thus connected to the form of the dispersion and the existence of a
non-zero sound velocity.

\subsection{Two dimensions}
\label{keeling_sec:two-dimensions}

In two dimensions, the distinction between superfluidity and BEC
becomes even more important, since an homogeneous two dimensional
system is unable to show true long-range order due to the
Mermin-Wagner theorem~\cite{mermin66}.  This can be seen by
considering the correlation function $\langle \psi^\dagger(\vec r)
\psi(\vec r^\prime) \rangle \simeq \rho_0 \langle \exp[i(\theta(\vec
r^\prime) - \theta(\vec r))] \rangle$.  Even for a system with a
non-zero phase stiffness, $E[\theta(\vec{r})] = (K_s/2) \int
d^2\vec{r} (\nabla \theta)^2$, the vanishing energy cost of long
wavelength phase twists leads to a thermal expectation which, at long
distances, takes the form
\begin{equation}
  \label{keeling_eq:3}
  \left< e^{i(\theta(\vec r^\prime) -\theta(\vec r))} \right>
  \propto 
  \exp\left( - \alpha_s \ln|\vec r - \vec r^\prime| \right) \propto |\vec r - \vec r^\prime|^{-\alpha_s}, \qquad
  \alpha_s=\frac{k_B T}{2 \pi K_s}\,,
\end{equation}
which vanishes algebraically as $|\vec r - \vec
r^\prime|\rightarrow\infty$.  Therefore, there is no ODLRO, and so no
single mode is macroscopically occupied; i.e., there is no
BEC. Nonetheless, superfluidity can survive, because the phase
stiffness $K_s$ implies a resistance to rotation of the low energy
modes of the condensate.  In fact, either by directly comparing the
calculation of phase stiffness and superfluid
density~\cite{griffin94}, or by calculating the current-current
response function from the parametrisation and energy functional
above~\cite{huang,pitaevskii03}, one finds that the phase stiffness
(and thus the power law decay) is directly related to the superfluid
stiffness: specifically, $K_s = \rho_s/ m^2$ where $m$ is the
quasiparticle mass.

In addition to the distinct nature of the low temperature phase in two
dimensions, the transition to this phase is also unusual.  As noted
above, a superfluid can be made to rotate by creating quantised
vortices, in which the phase winds by $2\pi$ around a point, and the
density of the condensate vanishes at that point.  As discussed by
Kosterlitz and Thouless~\cite{kosterlitz73,kosterlitz74} and
Berezhinskii~\cite{Berezinskii1972}, the transition out of the
superfluid phase occurs through the proliferation of these vortices.
The phase boundary can be found by a renormalisation   group
approach~\cite{kosterlitz74,nelson77,ChaikinLubensky:Book}, which
predicts that if $K_s < (2/\pi) k_BT$, vortices proliferate and
correlations decay exponentially, whereas for $K_s > (2/\pi) k_BT$,
vortices are irrelevant at large scales and correlations decay
algebraically. As a result, the exponent for the algebraic decay,
Eq.~(\ref{keeling_eq:3}) takes the universal value $\alpha_s =1/4$ at
the phase boundary, and the superfluid density undergoes a universal
jump~\cite{nelson77}.

\section{Modelling driven dissipative condensates}
\label{keeling_sec:driv-diss-syst}

A driven dissipative system is by definition one that is coupled to
more than one environment, or is driven by a time dependent pumping
field.  One can therefore no longer apply equilibrium concepts, such
as minimising free energy, or the fluctuation-dissipation theorem.
For a given system, such as a weakly interacting dilute Bose gas,
there are many different types of driving and dissipation.  Some of
these conserve particle number; others do not.  In this chapter we
consider the latter case, motivated by systems such as microcavity
polaritons (for a review see~\cite{Carusotto2013a} and chapters XX of
this book).

Microcavity polaritons are superpositions of excitons and photons, and
the photon part can leak out through the mirrors. To reach a steady
state this loss must be replenished by a compensating pump.  Several
models can be written to describe this, with varying descriptions of
the reservoir that replenishes the
condensate~\cite{wouters06,Szymanska2006,Szymanska2007}.  All such
models lead to the same essential differences between the equilibrium
and driven dissipative cases. We consider the model starting from the
weakly interacting dilute Bose gas,
\begin{displaymath}
  \hat H = \int d^d r 
  \hat\psi^\dagger(r)\left(-\frac{\nabla^2}{2m}\right) \hat\psi(r)
  +
  \frac{U}{2} \hat\psi^\dagger(r)\hat\psi^\dagger(r) \hat\psi(r)
  \hat\psi(r),
\end{displaymath}
and consider loss terms described by the quantum master equation
\begin{equation}
  \label{keeling_eq:5}
  \partial_t \rho = -i[\hat H, \rho]
  + \int d^dr \left(
    \frac{\kappa}{2} \mathcal{L}[\hat \psi(r),\rho]
    +
    \frac{\gamma}{2} \mathcal{L}[\hat \psi^\dagger(r),\rho]
    +
    \frac{\Gamma}{4} \mathcal{L}[\hat \psi^2(r),\rho]
  \right),
\end{equation}
where
$\mathcal{L}[\hat X,\rho] = 2 \hat{X} \rho \hat{X}^\dagger - [\hat{X}^\dagger
\hat{X}, \rho]_+$
is the usual Lindblad operator. The terms in Eq.~\eqref{keeling_eq:5} describe single
particle loss, single particle incoherent pump, and two-particle losses at rates
$\kappa,\gamma$, and $\Gamma$, respectively.

\subsection{Mean-field description \& collective excitations}
\label{keeling_sec:mean-field-descr}

The role of the dissipative terms in Eq.~(\ref{keeling_eq:5}) can be seen by
considering the corresponding mean-field equation of motion,
found by replacing $\langle \psi(\vec{r}) \rangle = \varphi(\vec{r})$, and
decoupling all correlators.  This gives:
\begin{equation}\label{keeling_eq:MF}
  i  \partial_t \varphi
  = \left[ -\frac{\nabla^2}{2m} + U|\varphi|^2 
    + \frac{i}{2} \left( \gamma- \kappa - \Gamma|\varphi|^2  \right)
  \right] \varphi,
\end{equation}
which is a modified Gross--Pitaevskii equation~\cite{Aranson2002}
(GPE) including dissipative terms describing particle gain and loss.
The nonlinear term with coefficient $\Gamma$ describes gain
saturation, or feedback between the condensate and the reservoir,
which prevents the particle density from diverging.  For small
fluctuations around the steady state, $\Gamma |\varphi_0|^2 =
\gamma_{\mathrm{net}} \equiv \gamma-\kappa$, one finds the
fluctuations have a complex spectrum $\omega_{\vec k}$ of the form:
\begin{equation}
  \label{keeling_eq:6}
  \omega_{\vec k} = - i \left(\frac{\gamma_{\mathrm{net}}}{2} \right)
  \pm \sqrt{\xi_{\vec k}^2 - \left(\frac{\gamma_{\mathrm{net}}}{2}\right)^2}, \qquad
  \xi_{\vec k}^2 = \frac{k^2}{2m} \left(\frac{k^2}{2m} 
    + 2 U |\varphi_0|^2 \right).
\end{equation}
The quantity $\xi_{\vec k}$ reduces to the equilibrium Bogoliubov
excitation spectrum in the limit $\kappa, \gamma, \Gamma \to 0$.  As
discussed in Sec.~\ref{keeling_sec:superfl-phase-coher}, $ \xi_{\vec
  k}$ has a linear dispersion, $\xi_{\vec k} \simeq c_s |\vec k|$ at
low momentum, where $c_s^2 = U |\varphi_0|^2 / m$. In contrast,
$\omega_{\vec k}$ is diffusive at low momentum, $\omega_{\vec k} = - i
D k^2$ with $D= c_s^2/ \gamma_{\mathrm{net}}$.

This modified dispersion would appear to mean that the critical
velocity vanishes.  Nonetheless, signatures of a non-vanishing
critical velocity can survive in static correlation functions ---
albeit washed out by dissipation. To see this one may calculate
theoretically the drag force on a defect immersed in a steady flow
pattern around the defect potential $V_\text{defect}(\vec r)$.  The
static drag force on the defect is given by~\cite{astrakharchik04}
$\vec{F}_{\text{drag}} \propto \int d^d \vec r \rho(\vec r) \nabla
V_\text{defect}(\vec r)$.  For a perfect superfluid below the critical
velocity, the flow pattern is symmetric, so that the effects of
pressure ahead of and behind the defect cancel: i.e.  ``d'Alembert's
paradox'' occurs, that an irrotational flow produces no drag.  The
GPE, with $\rho(r)=|\varphi(r)|^2$ describes such a perfect
superfluid, and would thus normally predict zero drag, however, if one
calculates the drag using the complex GPE given above, a finite drag
force exists at all velocities~\cite{Wouters2010,cancellieri_10}.
There does however remain a marked threshold at $v=c_s$ above which
the drag increases more rapidly with velocity.  
This behaviour is
similar to an equilibrium superfluid at finite temperature, however
calculating the  finite temperature drag for an equilibrium
superfluid requires including fluctuations beyond the mean-field
theory.

The modified dispersion also has effects on the linear-response
calculation of phase correlations~\cite{Roumpos2012,Chiocchetta2013},
however as discussed below, a more dramatic change arises because a
linearised theory becomes inadequate in calculating correlations of
the driven dissipative Bose gas in two dimensions.

\subsection{Beyond mean-field description}
\label{keeling_sec:schw-keldysh-acti}

In order to calculate correlations and response functions, the
mean-field description is not sufficient.  Various methods for dealing
with driven dissipative systems such as Eq.~(\ref{keeling_eq:5}) exist.  We
consider an approach starting from the Schwinger-Keldysh path integral
(see Ref.~\cite{Kamenev2011} for an introduction), defined by the
``partition sum''
\begin{equation}\label{keeling_eq:Z}
\mathcal{Z} = \int \mathcal D (\psi_C,\psi_Q) e^{i S[\psi_C,\psi_Q]},
\end{equation}
where $S$ is the  Schwinger-Keldysh action
written in terms of ``classical'' and ``quantum'' fields $\psi_C, \psi_Q$.
For the model in Eq.~(\ref{keeling_eq:5}) this takes the form:
\begin{multline}
  \label{keeling_eq:7}
  S[\psi_C, \psi_Q] = 
  \int dt d^dr \Big\{
  \begin{pmatrix}
    \bar{\psi}_C & \bar{\psi}_Q
  \end{pmatrix}
  \begin{pmatrix}
    0 & [D^A_0]^{-1} \\
    [D^R_0]^{-1}  & [D^{-1}_0]_K
  \end{pmatrix}
  \begin{pmatrix}
    \psi_C \\ \psi_Q
  \end{pmatrix}
  \\
 - \left[
    \left( \tfrac{U}{2} + i \tfrac{\Gamma}{4} \right)
    \left( (\bar{\psi}_C^{2} + \bar{\psi}_Q^{2}) \psi^{}_C\psi^{}_Q \right)
  + \text{c.c.} \right] + i \Gamma  \bar\psi^{}_C\psi^{}_C\bar\psi^{}_Q\psi^{}_Q \Big\},
\end{multline}
where the bare inverse retarded Green's function is given by
$[D^R_0]^{-1} = i\partial_t - [-\nabla^2/(2m) + i (\gamma - \kappa)/2
]$ in time and real space domain, or $\omega - [\epsilon_k + i (\gamma
- \kappa)/2]$ in the frequency and momentum domain. The inverse
advanced Green's function $[D^A_0]^{-1}$ is the complex conjugate of
this, and the Keldysh component of the inverse bare Green's function
$[D^{-1}_0]_K = i(\kappa+\gamma)$ describes the noise associated with
both pumping and decay.  Despite the existence of terms which create
and destroy particles, Eq.~(\ref{keeling_eq:7}) still possesses a
$U(1)$ symmetry under simultaneous phase rotations of the fields
$\psi_C, \psi_Q$; as such there still exists the possibility of
spontaneous symmetry breaking, phase stiffness, and of superfluidity.

Equations~(\ref{keeling_eq:Z},\ref{keeling_eq:7}) are fully equivalent to the quantum
master equation, Eq.~\eqref{keeling_eq:5}, but written in a functional
integral formulation. This formulation allows one to apply a wide
range of techniques from quantum field theory.  A first, rather generic
simplification consists in taking the \emph{semiclassical limit},
which can be justified by a power counting argument. This is strictly
justified close to the condensation threshold, where $\gamma - \kappa
\to 0$, but provides a useful approximation also away from this 
limit. At threshold, the retarded and advanced inverse Green's
functions scale as $\sim k^2$ with $\omega \sim k^2$, while there is
no scaling of the Keldysh component, $[D^{-1}_0]_K = i(\kappa+\gamma)
\sim k^0$. Using these, along with the natural scaling $dr \sim
k^{-1}, dt \sim \omega^{-1}$, we can then determine the scaling
dimensions of the fields $\psi_C \sim k^{(d-2)/2}, \psi_Q \sim
k^{(d+2)/2}$ required in order that the quadratic contributions to the
action are dimensionless. This then allows determination of the
scaling of the various quartic terms.  Due to this scaling, any quartic
term involving more than a single quantum field is irrelevant --- i.e.
such terms scale to zero at long wavelength $k\to 0$, and can thus be
omitted in the semiclassical limit.  This provides direct contact to
the dissipative GPE~\eqref{keeling_eq:MF}.  The field equation obtained from
the saddle point, $\delta S /\delta \bar\psi_Q =0$, of Eq.~\eqref{keeling_eq:7}
in the semiclassical limit reads
\begin{equation}\label{keeling_eq:var} i \partial_t \psi_C = \left[
    -\frac{\nabla^2}{2m} + \frac{U}{2}|\psi_C|^2 + \frac{i}{2} \left( \gamma- \kappa -
      \frac{\Gamma}{2}|\psi_C|^2 \right) \right] \psi_C + i(\kappa+\gamma) \psi_Q.
\end{equation}
This almost matches Eq.~\eqref{keeling_eq:MF} if one identifies $\varphi =
\psi_C/\sqrt{2}$, but with an extra term involving the quantum field
$\psi_Q$.

While the ``classical'' field can acquire a finite expectation in the
condensed state, the ``quantum'' field has to vanish on average and
describes the noise. As a saddle point equation,
Eq.~\eqref{keeling_eq:var} neglects the fluctuations of the fields
$\psi_C, \psi_Q$, appearing in the full functional integral.
Remarkably, this equation can be upgraded to a full description of the
problem by means of the Martin-Siggia-Rose
construction~\cite{Kamenev2011}.  This shows that the functional
integral Eq.~\eqref{keeling_eq:Z} in the semiclassical limit is
\emph{equivalent} to a stochastic partial differential equation.  In
our case, this is the driven dissipative stochastic Gross-Pitaevskii
equation (DSGPE) which is equivalent to Eq.~(\ref{keeling_eq:var})
with the replacement $i (\kappa + \gamma) \psi_Q \to \xi(\vec r, t)$,
where $\xi(\vec r, t)$ describes a Gaussian white noise process
characterised by $\langle \xi(\vec r, t) \rangle = 0$ and $\langle
\xi(\vec r, t) \bar\xi(\mathbf{r}', t') \rangle =
\frac{\gamma+\kappa}{2} \delta(t - t') \delta(\mathbf{r} -
\mathbf{r}')$, and vanishing off-diagonal correlators. In this sense,
the DSGPE corresponds to a fully fluctuating (semiclassical) many-body
problem --- in stark contrast to the deterministic GPE
\eqref{keeling_eq:MF}.

\subsection{Equilibrium vs. non-equilibrium dynamics}
\label{keeling_sec:noneq}

The action~\eqref{keeling_eq:7} in the semiclassical limit or the DSGPE allow
us to state precisely the sense in which the driven dissipative system
is a genuinely non-equilibrium situation.  To this end we rewrite
Eq.~\eqref{keeling_eq:var} by splitting the deterministic parts on the RHS
into coherent (reversible) and dissipative (irreversible)
contributions  labelled $c, d$.  To avoid confusion with these
labels we omit the suffix $C$ on the classical field $\psi_C$.  The
DSGPE takes the form:
\begin{equation}
  \label{keeling_eq:12}
  i  \partial_t \psi = \frac{\delta H_c}{\delta\bar\psi} - i  \frac{\delta H_d}{\delta\bar\psi}  + \xi
\end{equation}
with effective coherent and dissipative Hamiltonians
\begin{equation}
  \label{keeling_eq:11}
  H_{\alpha=c,d} = \int d^d r \left( K_\alpha |\nabla\psi|^2 + r_\alpha
    |\psi|^2 + \frac{u_\alpha}{2} |\psi|^4 \right).
\end{equation}
The coefficients are $K_c = 1/2m, K_d =0, r_c =0, r_d =
(\gamma-\kappa)/2, u_c = U/2$ in our problem, and $u_d =
\Gamma/4$. Note, however, that the value of $r_c$ is adjustable by a
gauge transformation $\psi \mapsto \psi e^{-i \omega t}$ such that
$r_c \mapsto r_c - \omega$. It can be
shown~\cite{Graham1990,Sieberer2014} that if the system is to relax to
thermodynamic equilibrium (if the steady state of
Eq. (\ref{keeling_eq:12}) is to be described by a Gibbs distribution),
then the coherent and dissipative Hamiltonians must be proportional to
each other; that is,
\begin{equation}
  \label{keeling_eq:16}
  H_c = \nu H_d \quad \Leftrightarrow \quad  \nu = \frac{K_c}{K_d} = \frac{u_c}{u_d}
\end{equation}
where $\nu$ is a constant.

This requirement is in general not satisfied for a driven dissipative
system because the microscopic origins of reversible and irreversible
dynamics are independent. For example, in the microscopic description
of Eq.~\eqref{keeling_eq:5} the rates can be tuned independently from
the Hamiltonian parameters, as they have completely different physical
origins.

We have thus far made two important observations pertaining to a
driven condensate. One is that, in spite of particle number
non-conservation, the equations of motion describing an incoherently
driven condensate enjoy a $U(1)$ phase symmetry.  The second
observation is that the steady state cannot be in general described by
an equilibrium ensemble. So, while a condensation transition involving
spontaneous symmetry breaking is possible, the nature of the
transition and the conditions under-which such a condensate would be
indeed a stable fixed point of the dynamics may be different from the
equilibrium case.

The question concerning the nature of the condensation transition in
three dimensional driven condensates was addressed in
Refs.~\cite{Sieberer2013,Sieberer2014,Tauber2014}.  One main result of
this analysis was that the equilibrium symmetry, $H_c=\nu H_d$ is
emergent in the low frequency limit even though it is not present
microscopically. Hence the \emph{correlation} functions correspond to
an effective equilibrium description with a well defined emergent
temperature. Nevertheless, the drive conditions affect the long
wavelength dynamical \emph{response} functions, guaranteed by the
existence of a new and independent critical exponent.

In two dimensions, as we will show next, the long wavelength behaviour
is changed much more dramatically: in isotropic systems, the slow
algebraic decay of correlation functions that occurs in equilibrium is
replaced by far faster exponential or stretched exponential
decay. Only for sufficiently anisotropic systems can the
quasi-long-ranged algebraic decay found in equilibrium be recovered.

\section{Long wavelength fluctuations and phase coherence}
\label{keeling_sec:phase-coher-driv}

As discussed in Sec.~\ref{keeling_sec:superfl-phase-coher}, the
Mermin-Wagner theorem states that a homogeneous equilibrium system
with a continuous symmetry cannot show long-range order that breaks
that continuous symmetry in two or fewer dimensions. Rather, in two
dimensions, long-wavelength phase fluctuations lead to algebraic decay
of order parameter correlations. Naturally the question arises whether
this statement remains true out of equilibrium.  The answer to this
question proves to depend on both dimensionality, and anisotropy.  In
three dimensions, the deviation from effective equilibrium, which is
encoded in the difference between the ratios $K_c/K_d$ and $u_c/u_d$
(cf.\ Eq.~\eqref{keeling_eq:16}), vanishes in the long-wavelength
limit, both close to criticality~\cite{Sieberer2013}, and in the
ordered (Bose condensed) phase~\cite{Altman2015}. However, in
isotropic two dimensional systems, these non-equilibrium effects are
relevant~\cite{Altman2015}, and ultimately lead to the destruction of
algebraic order. There is, however, a loophole: a spatially
\emph{anisotropic} system can support an algebraically ordered
phase. These conclusions follow from a hydrodynamic description of the
order parameter dynamics, which we review in the following.

In order to allow for spatial anisotropy, we consider a generalisation
of the model described by Eq.~\eqref{keeling_eq:12}, in which the
gradient terms in the effective Hamiltonians~\eqref{keeling_eq:11} are
replaced by $\sum_{i = x,y} K_{\alpha}^i |\partial_i \psi|^2$. As
described in Sec.~\ref{keeling_sec:superfl-phase-coher}, a
hydrodynamic description can be obtained by employing a density-phase
representation, $\psi = \sqrt{{\rho}} e^{i \theta}$, leading to
coupled equations of motion for the density $\rho$ and phase
$\theta$. Eliminating the gapped fluctuations of the density around
its mean value $\rho_0$, and keeping only the leading terms in a
low-frequency and low-momentum expansion in the remaining equation for
the phase, we obtain the anisotropic
Kardar-Parisi-Zhang~\cite{Kardar1986} (KPZ)
equation~\cite{Altman2015},
\begin{equation}
  \label{keeling_eq:13}
  \partial_t \theta = \sum_{i = x, y} \left[ D_i \partial_i^2 \theta +
    \frac{\lambda_i}{2} \left( \partial_i \theta \right)^2 \right] + \eta,
\end{equation}
where $\eta(\vec r, t)$ is a Gaussian stochastic noise with zero mean,
$\langle \eta(\vec r, t) \rangle = 0$, and second moment $\langle
\eta(\vec r, t) \eta(\vec r', t') \rangle = 2 \Delta \delta(\vec r -
\vec r')\delta(t- t')$, with
\begin{equation}
  \label{keeling_eq:14}
  \Delta = \frac{\left( \kappa + \gamma \right) \left( u_c^2 + u_d^2 \right)}{2
    u_d \left( \kappa - \gamma \right)}.
\end{equation}
The effective diffusion constants in Eq.~\eqref{keeling_eq:13} and the non-linear
couplings are given by
\begin{equation}
  \label{keeling_eq:15}
  D_i  = K_c^i \left( \frac{K_d^i}{K_c^i} + \frac{u_c}{u_d} \right), \qquad
  \lambda_i  = -2 K_d^i \left( \frac{K_c^i}{K_d^i} - \frac{u_c}{u_d} \right).
\end{equation}
Evidently the non-linear terms in the KPZ equation vanish when the
equilibrium condition $K_c^x/K_d^x = K_c^y/K_d^y = u_c/u_d$, which
generalises Eq.~\eqref{keeling_eq:16} to the spatially anisotropic case, is
met. The degree of anisotropy is measured by the anisotropy parameter
$\Phi = \lambda_y D_x/\lambda_x D_y$: when $\Phi \neq 1$, the system
is anisotropic.

An important difference exists between our KPZ model and the original
context of this equation, as an equation for the interface height in a
model of randomly growing interfaces~\cite{Kardar1986}.  The analogue
of the interface height in our model is actually a phase, $\theta$,
and the phase is compact, i.e. $\theta \equiv \theta + 2\pi$.  This
means that topological defects in this field --- vortices --- are
possible.  This difference with the conventional KPZ equation also
arises in ``Active Smectics''~\cite{Chen2013}.  Analysis of
Eq.~(\ref{keeling_eq:13}) in the absence of vortices is the analogue
of the low temperature spin-wave (linear phase fluctuation) theory of
the equilibrium $XY$ model.  Indeed, without the non-linear terms, the
KPZ equation reduces to linear diffusion, which would bring the field
to an effective thermal equilibrium with power-law off-diagonal
correlations (in $d=2$). A transition to the disordered phase in this
equilibrium situation can occur only as a Kosterlitz-Thouless (KT)
transition through proliferation of topological defects in the phase
field.

Our aim is to obtain the behaviour of correlations of the condensate
field $\psi$ at large distances.  We have taken the first step of
reducing this task to finding the correlations of the phase field
$\theta$, whose dynamics are given by the KPZ
equation~\eqref{keeling_eq:13}. But this equation contains more
information than is actually required: in particular, it involves
fluctuations with all wavelengths, ranging from the microscopic
condensate healing length $\xi_0$ up to the largest scales of order of
the linear system size $L$ (we consider a square 2D system of area $L^2$).
Because of the nonlinear term, fluctuations at different wavelengths
couple to each other.  Our goal, therefore, is to eliminate the short
scale fluctuations, and in this way obtain an effective description of
the system at large scales.

This idea is implemented by the renormalisation group
procedure~\cite{ChaikinLubensky:Book}. To this end, we decompose the
phase field into short- and long-wavelength components, above and
below some length scale $\ell$.  We then integrate out the
short-wavelength components, treating the nonlinear terms
$\lambda_{x,y}$ which couple short- and long-wavelength components
perturbatively.  As such this approach is limited to
close-to-equilibrium conditions in which the couplings $\lambda_{x,y}$
are small%
\endnote{More precisely, the perturbation theory is valid when a
  suitable dimensionless measure of the ratio of the non-linear
  $\lambda_{x,y}$ terms in the KPZ equation to the linear ones is
  small; this measure proves to be the parameter $g$ defined below.}.
This procedure is then iterated for increasing lengthscales $\ell$,
successively integrating out the short wavelength components.  In real
space this corresponds to repeated coarse-graining, eliminating fine
details on scales shorter than $\ell$.  Performing this program for
the anisotropic KPZ equation~\cite{Wolf1991,Chen2013}, the resulting
equation for the long-scale components of the phase field again takes
the form~\eqref{keeling_eq:13}, but with coefficients that are
modified by the short-scale fluctuations.

\begin{figure}
  \centering
  \includegraphics[width=3.5in]{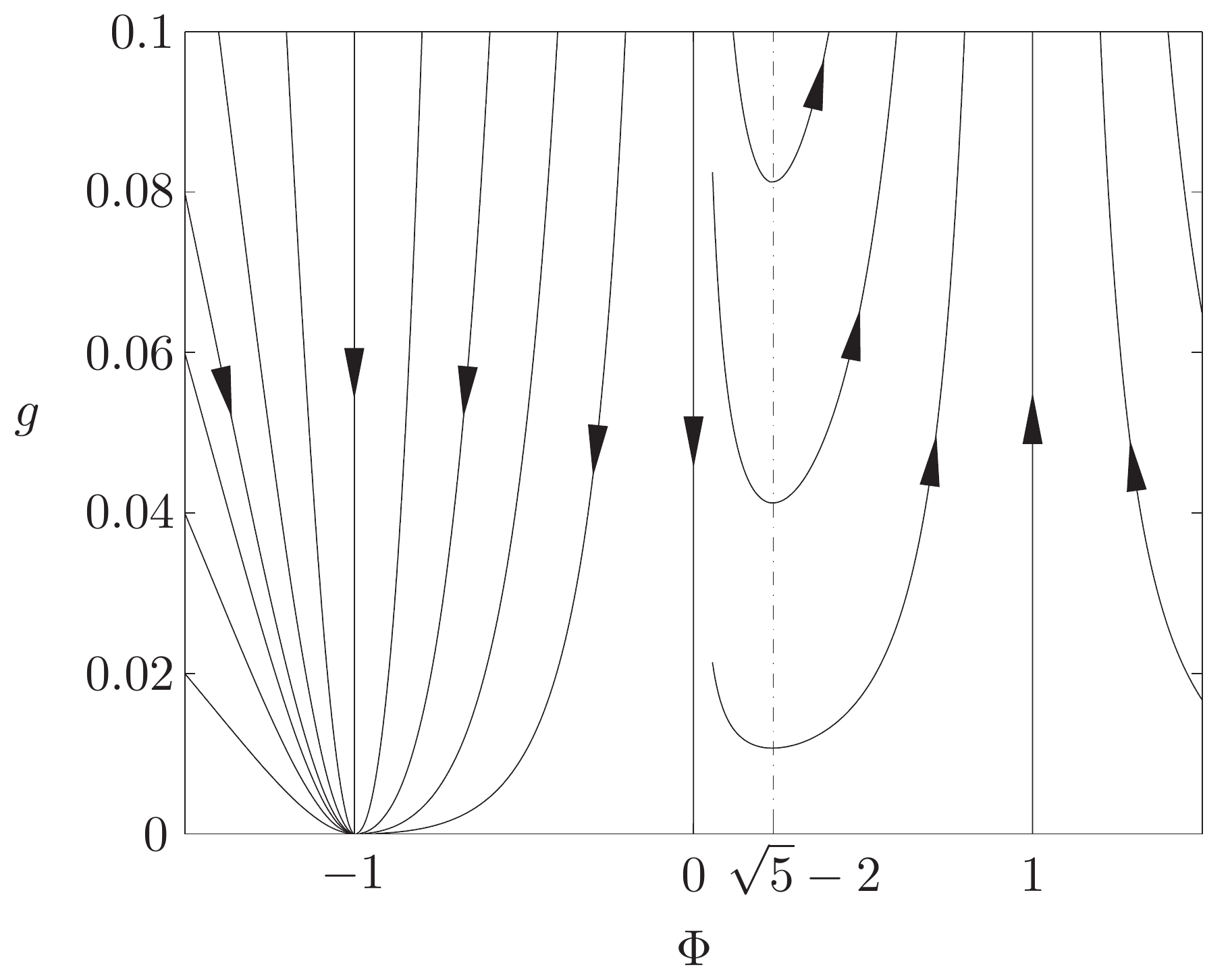}
  \caption{RG flow for the anisotropic KPZ equation in 2D. In the
    weakly anisotropic regime $\Phi > 0$, the flow lines approach the
    isotropic line $\Phi = 1$ and flow towards $g \to \infty$; for
    $\Phi < 0$, in the strongly anisotropic regime, they converge to a
    stable effective equilibrium fixed point with $\Phi = -1$ and $g =
    0$.}
  \label{keeling_fig:aniso_KPZ_flow}
\end{figure}

The modification of the coefficients as one goes to long length scales
can be characterised entirely by the ``flow'' of two dimensionless
parameters: These are a dimensionless form of the non-linearity, $g
\equiv \lambda_x^2 \Delta/D_x^2 \sqrt{D_x D_y}$ and the anisotropy
parameter $\Phi = \lambda_y D_x/\lambda_x D_y$ introduced below
Eq.~\eqref{keeling_eq:15}.  The parameter $g$ describes the importance
of the nonlinear terms $\lambda_x$, for ``typical'' (i.e. root mean
squared) fluctuations of the field $\theta$.  The mean squared
fluctuations of $\theta$, according to the linear theory, are
proportional to the noise strength $\Delta$, which drives the
fluctuations, and inversely proportional to the geometric mean of the
diffusion coefficients $D_{x,y}$, which smooth out those fluctuations.
Knowing the anisotropy parameter $\Phi$, together with $g$, then
allows us to estimate the importance of the other non-linearity
$\lambda_y$.  The microscopic parameters of the system determine the
``bare'' values $g_0$ and $\Phi_0$ at the starting length scale $\ell
= \xi_0$.  The values of $g, \Phi$ at some other scale $\ell > \xi_0$
are obtained by integrating the RG flow equations
\begin{equation}
  \label{keeling_eq:17}
  \begin{split}
    \frac{d g}{d l} & = \frac{g^2}{32 \pi} \left( \Phi^2 + 4 \Phi - 1
    \right), \\ \frac{d \Phi}{d l} & = \frac{\Phi g}{32 \pi} \left( 1 -
      \Phi^2 \right),
  \end{split}
\end{equation}
with the logarithmic scale $l = \ln(\ell/L)$. The resulting RG flow is
illustrated in Fig.~\ref{keeling_fig:aniso_KPZ_flow}.
There are two distinct flow patterns to the left and right of the line
$\Phi = 0$ and we discuss these in turn.

For $\Phi > 0$ the flow with increasing $\ell$ is towards strong
coupling, $g \to \infty$, and isotropy, $\Phi \to 1$.  Thus, in this
regime, which we denote \emph{weak anisotropy}, the approximation of
treating $g$ perturbatively eventually fails.  Simulations of the
isotropic KPZ
equation~\cite{Kim1989,Miranda2008,Marinari2000,Ghaisas2006,Chin1999,Tang1992}
find that correlations of the phase field $\langle (\theta(\vec r) -
\theta({\vec r}^\prime))^2\rangle$ scale algebraically with separation
$|\vec{r}-\vec{r}^\prime|$. This suggests that ultimately the flow of
$g$ must terminate at a strong coupling fixed point $g_{*}$, which is
however beyond the scope of the perturbatively derived flow
equations~\eqref{keeling_eq:17}.  This scaling of $\langle
(\theta(\vec r) - \theta({\vec r}^\prime))^2\rangle$ would lead,
according to Eq.~\eqref{keeling_eq:3} to stretched exponential decay
of condensate field correlations.

Note, however, that in our analysis of the KPZ
equation~\eqref{keeling_eq:13} we have so far neglected topological
defects, which can exist in the compact field $\theta$.  Proliferation
of such defects at the strong coupling fixed point would lead to
exponential (i.e., even faster) decay of correlations of $\psi$.

The regime of \emph{strong anisotropy}, on the other hand, corresponds
to the region $\Phi < 0$. There the flow lines terminate for $\ell \to
\infty$ in an effective \emph{equilibrium} fixed point with $g = 0$
and $\Phi = -1$.  Thus, in this regime, the effective description of
the system at large scales approaches the equilibrium description
(note that $g \propto \lambda_x^2$ and hence $g = 0$ in equilibrium),
and so algebraic correlations of the condensate field can survive (as
long as one is below the KT transition temperature).

The possibility of a flow to strong coupling is in stark contrast to
the 3D case in which even in isotropic systems with $\Phi = 1$ small
deviations $g \ll 1$ from equilibrium are irrelevant and flow to zero
as $\ell \to \infty$, leading to the effective equilibrium physics
discussed at the end of Sec.~\ref{keeling_sec:noneq}.  Note, however,
that even in 3D, for sufficiently strong drive and dissipation, i.e.,
values of $g$ larger than some critical value, there may be a
non-equilibrium transition to the disordered phase, described by the
strong coupling fixed point of the 3D KPZ equation~\cite{Fisher1992}.
In one dimension, even equilibrium systems show only short range
(exponential) correlations at non-zero temperatures.  Nonetheless, it
is still possible to see the effect of the KPZ nonlinearity on the
scaling of the spatial~\cite{Gladilin2013} and temporal
coherence~\cite{Kai15,He2014} of a one dimensional condensate.

\subsection{Current experiments, weak anisotropy, BKT physics and crossovers}
\label{keeling_sec:exper-prob-coher}

As discussed in chapter 25 [Kim, Nitsche, Yamamoto], experiments on
incoherently pumped polariton
condensates~\cite{Roumpos2012,Nitsche2014} have measured an apparent
algebraic decay of correlations, by measuring the fringe visibility in
an interference experiment.  Similar results have also been seen in
numerical experiments on a parametrically pumped (OPO)
system~\cite{Dagvadorj2014}.  Since, as discussed above, algebraic
order is destroyed at large scales, a question arises about the
interpretation of these experiments.  In this section we discuss how,
although the asymptotic behaviour at $\ell \to \infty$ does lead to
the strong coupling fixed point, the length scales $\ell$ required to
see this may be very large, particularly when the condensate is well
developed.

Current experiments with exciton-polaritons fall into the regime of
weak anisotropy.  This anisotropy results from the interplay between
polarisation pinning to the crystal structure, and the splitting of
transverse electric and transverse magnetic cavity
modes~\cite{Carusotto2013a,Shelykh2010} --- taken together these mean
that there is anisotropy between directions parallel and perpendicular
to this pinned lattice direction.  As discussed above, for weak
anisotropy, the flow is to strong coupling and algebraic order is
absent on the largest scales.  However at intermediate scales,
$g(\ell) \ll 1$, and so correlations can still decay algebraically. It
is therefore natural to ask how large the system must be to see the
breakdown of algebraic correlations --- i.e. what system size $L$ is
required such that $g(L) \simeq 1$.  Setting $\Phi = 1$ in the flow
equation for $g$ in Eq.~\eqref{keeling_eq:17}, we find that if the
microscopic parameters in Eq.~\eqref{keeling_eq:11} correspond to
close-to-equilibrium conditions, i.e., if the bare value $g_0 \ll 1$,
then the renormalised value $g = 1$ is reached only at the
exponentially large characteristic KPZ scale $L=L_{*} = \xi_0 \exp(8
\pi/g_0)$. Starting from a microscopic model for polariton
condensation~\cite{Carusotto2013a} that models the excitonic reservoir
explicitly, and which provides a more faithful description of the
condensation dynamics than the model of Eq.~\eqref{keeling_eq:5}, we
obtain the expression for the bare non-linearity~\cite{Altman2015}
\begin{equation}
  \label{keeling_eq:19}
  g_0 = \frac{2 u_c \bar{\gamma}^2}{K_c} \frac{\bar{\gamma}^2 + \left( 1 + x
    \right)^2}{x \left( 1 + x \right)^3},
\end{equation}
which depends on the dimensionless net pumping rate $x = \gamma/\kappa
- 1,$ and the dimensionless combination $\bar{\gamma} = \kappa
R/\gamma_R u_c,$  where $R$ and $\gamma_R$ are the rate of
scattering between the excitonic reservoir and condensate, and the
reservoir relaxation rate, respectively.  The dimensionless parameter
$x$ gives a measure of how far above ``threshold'' the system is
pumped.  For high pump rates the KPZ scale $L_{*}$ grows rapidly
beyond any reasonable system size, so that a sufficiently strongly
pumped system will always appear algebraically ordered up to system
size $L$.  Such a system resides effectively in equilibrium, with a
temperature set by the noise strength.  The form of
Eq.~(\ref{keeling_eq:19}) implies that for \emph{any} finite system
size $L$, it is always possible to choose a value $x$ large enough
that algebraic correlations are seen over the entire system.  As such,
the experimental observation of algebraic correlations is perfectly
consistent with the results here.

At weak pump rates, near the threshold $x \to 0$, the bare coupling
$g_0$ grows and so the critical size $L_\ast$ decreases, and may
become comparable to the system size $L$.  This then prompts a second
question: as the pump strength is reduced, does the algebraic order
break down before the KT transition occurs?  As long as $L_\ast \gg
L$, the system is effectively thermal and consequently it can undergo
a KT transition to a disordered phase as described in
Sec.~\ref{keeling_sec:superfl-phase-coher} if $x$ is decreased below a
critical value $x_{\mathrm{KT}}$.  One can then evaluate the length
scale $L_\ast$ at this pump rate, i.e. we denote $L^\prime \equiv
L_\ast(x=x_{KT}) \approx \xi_0 \exp(2/\bar{\gamma}^2)$.  If the system
is much smaller than this critical size (i.e., $L\ll L^\prime$), then
even at the KT point, $L\ll L_\ast,$ and so the KPZ physics does not
become apparent. This is consistent with recent experiments on exciton
polaritons~\cite{Nitsche2014} which showed the behaviour expected of a
KT transition as discussed in Sec.~\ref{keeling_sec:two-dimensions}
--- i.e. a transition between algebraic order with exponent $\alpha_s=
1/4$ right at the transition and short ranged, exponentially decaying
order.  Such behaviour is consistent with a system of size $L\ll
L^\prime$.  However, if the system is sufficiently
large~\cite{Altman2015}, i.e., $L \gg L^\prime$, then algebraic order
at length scale $L$ will break down before the KT transition.  The
dependence of $L^\prime$ on $\bar{\gamma}$ can be understood
intuitively: as $\bar{\gamma} \to 0$, then the polariton lifetime
diverges, and thermalisation is perfect for any finite size
system. Therefore $L^\prime \to \infty$ and strong coupling with $g
\approx 1$ is never reached.  Thus, in order to clearly see the
breakdown of algebraic order, one should increase the polariton loss
rate $\kappa$.

\subsection{Strong anisotropy, re-entrant phase transition}
\label{keeling_sec:strong-anisotropy-re}

While the above analysis shows that the algebraic order observed in
recent experiments with (nearly) isotropic polariton systems~(see
chapter 25 [Kim, Nitsche, Yamamoto] and
Refs.~\cite{Roumpos2012,Nitsche2014}) must be an intermediate scale
crossover phenomenon, true algebraic order in the thermodynamic limit
is nevertheless possible in the \emph{strong anisotropy} regime,
$\Phi_0 < 0$ as discussed above. We discuss here the experimental
consequences this would have for a sufficiently anisotropic polariton
system.

\begin{figure}
  \centering
  \includegraphics[width=3.3in]{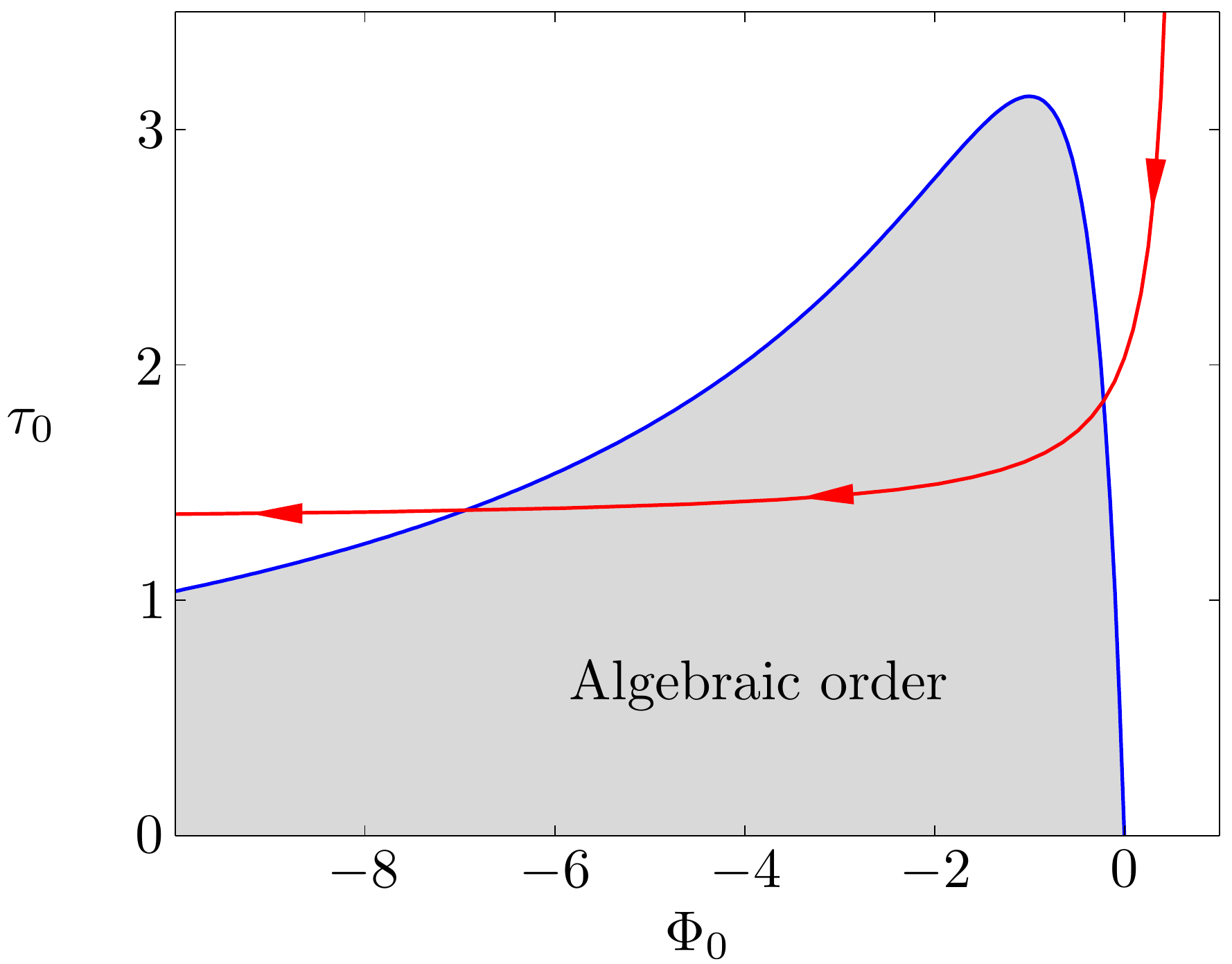}
  \caption{Phase diagram of a 2D anisotropic driven dissipative system
    for low noise. The thin line corresponds to values of $\Phi_0$ and
    $\tau_0$ derived from a microscopic model for polariton
    condensation~\cite{Altman2015}, with the arrows indicating the
    direction of increasing pump rate $\gamma$. Note that the ordered
    phase is first entered and then left again as $\gamma$ is
    increased. Values for the other dimensionless parameters are
    $\bar{\gamma} = 2, \bar{u} = 2, K_d^x/K_c^x = 1/8,$ and
    $K_d^y/K_c^y = 1/4$. }
  \label{keeling_fig:reentrance}
\end{figure}

The effective temperature of the system at the strong anisotropy fixed
point is given by the \emph{renormalised} value of the dimensionless
noise $\tau \equiv \Delta/\sqrt{D_x D_y}$.  Because the phase is a
compact variable, algebraic order only exists if this effective noise
temperature is low enough.  Specifically the KT transition occurs if
the renormalised value of this dimensionless noise reaches $\pi$.  One
may then derive a phase diagram by identifying the location of this condition
$\tau(\ell \to \infty)=\pi$ is in the manifold of \emph{bare}
couplings, i.e.  in terms of the microscopic experimental
parameters. To do this we must complement the RG flow
equations~\eqref{keeling_eq:17} for $g$ and $\Phi$ by additional
equations for $\tau$ and $D_{x,y}$ (see Ref.~\cite{Chen2013} for
details), and follow the flow to the effective equilibrium regime $g
\approx 0$.  We must also assume that, up to the length scale at which
the latter regime is reached, vortices are sufficiently  dilute that
their influence on the RG flows can be neglected. It turns out that
the renormalised value $\tau$ of the dimensionless noise strength
crosses the critical value for the KT transition for bare values
$\tau_0$ and $\Phi_0$ that are located on the curve determined by
\begin{equation}
  \label{keeling_eq:18}
  \tau_0 = - \frac{4 \pi \Phi_0}{\left( 1 - \Phi_0 \right)^2}.
\end{equation}
The resulting phase diagram of strongly anisotropic driven dissipative
systems in the $\Phi_0-\tau_0$ plane is depicted in
Fig.~\ref{keeling_fig:reentrance}.

It is interesting to work out the bare parameters $\Phi_0$ and
$\tau_0$ starting from the same microscopic model for polariton
condensation that led to the estimate of $g_0$ in
Eq.~\eqref{keeling_eq:19}.  This leads to a particular trajectory
through the $\Phi_0 - \tau_0$ phase diagram as one increases the
microscopic pump rate $\gamma$; this trajectory is shown on
Fig.~\ref{keeling_fig:reentrance}. An initially moderate anisotropy,
i.e., $\Phi_0 > 0$ but different from 1, becomes more substantial as
the pump rate is increased, so that $\Phi_0$ first becomes negative
and then, if the value of the dimensionless interaction strength
$\bar{u} = u_c/\sqrt{K_c^x K_c^y}$ is sufficiently small (for details
see Ref.~\cite{Altman2015}), crosses the boundary to the algebraically
ordered phase. Remarkably, upon pumping the system at an even higher
rate the ordered phase is then left again; that is, the transition is
{\it reentrant}.

\section{How to define and measure superfluidity in a dissipative
  system}
\label{keeling_sec:superfl-resp-funct}

In the previous section we have discussed how the presence of drive
and dissipation affect the low energy effective theory of the
two-dimensional system, and how these changes are manifested in the
correlation functions of the system.  In this section, we focus
instead on the current-current response function, $\chi_{ij}(\vec{q},
\omega)$ for a dissipative system, and how to identify whether a
superfluid fraction survives.  The discussion in this section is
focused on the case where phase fluctuations remain small, i.e. when
the nonlinear term in the KPZ equation does not dominate the physics.
As discussed above, this requires either a small finite system (as in
the current experiments), or a system with sufficiently anisotropic
interactions.  While the calculations presented below can easily be
extended to the anisotropic case, we present results only for the
isotropic situation.  When the nonlinearity in the KPZ equation
becomes large, and algebraic correlations are destroyed, there may
still exist a finite superfluid fraction; for a discussion of this
point see~\cite{Sieberer2015}.  However, if the growth of the KPZ
nonlinearity ultimately leads to vortex proliferation and short range
order, no superfluid fraction is expected to survive in the
thermodynamic limit.

The response function $\chi_{ij}(\vec{q},\omega=0)$ for a
non-equilibrium system can be calculated by defining a generating
functional for correlations of currents.  As noted in
section~\ref{keeling_sec:schw-keldysh-acti}, the driven dissipative
system no longer has a conserved current, but does still show a $U(1)$
phase symmetry.
  
How then does the system respond to a phase twist between the
boundaries of the system~\cite{Janot2013}?  Such a physical phase
twist couples to the unphysical ``quantum'' current $j_{Q,i}({\vec q})
= \sum_{\vec k} \left[ \bar{\psi}_{C,\vec{k+q}} \psi_{Q,\vec{k}}^{}
  + \bar{\psi}_{Q,\vec{k+q}} \psi_{C,\vec{k}}^{} \right]
\gamma_i(2{\vec k} + {\vec q})$.  To measure a response function we
must see how some physical quantity responds to such a phase twist.
For this we measure the standard particle current.  This leads us to
the generating functional:
\begin{align}
  \label{keeling_eq:8}
  \mathcal{Z}[\vec{f}, \vec{g}]
  &=
  \int \mathcal{D}(\bar \psi, \psi) \exp\left(
    i S[\bar \psi, \psi]
    + 
    i  S_j[\bar \psi, \psi] 
  \right), \\
  S_j[\bar \psi, \psi]  &=
  \sum_{\vec k, \vec q}
  \bar{\Psi}^T_{\vec{k+q}}
  \begin{pmatrix}
    g_i(\vec q) & f_i(\vec q) + g_i(\vec q) \\
    f_i(\vec q) - g_i(\vec q) & - g_i(\vec q)
  \end{pmatrix}
  \Psi_{\vec k} \;
  \gamma_i(2\vec k + \vec q),
  \nonumber
\end{align}
where $\Psi^T=(\psi_C \ \psi_Q)$ and $S[\bar \psi, \psi]$ is the
Schwinger-Keldysh action, e.g., Eq.~(\ref{keeling_eq:7})\endnote{In
  Ref.~\cite{Keeling2011} a different model action was taken,
  involving frequency dependent gain.  This difference does not affect
  the general points discussed below, but does affect several details
  of the calculation.}.  The source field $\vec f$ corresponds to the
phase twist, and couples to the quantum current.  The field $\vec g$
couples to the observable particle current; the strange form in
Keldysh space occurs in order to calculate normal-ordered expectations
from the Schwinger-Keldysh path integral\endnote{One can alternatively
  use a simpler form at the expense of introducing causality factors
  to ensure normal ordering.}.  Taking derivatives with respect to
these fields we find the current-current correlation function:
\begin{equation}
  \label{keeling_eq:9}
  \chi_{ij}(\vec q, \omega=0)
  = 
  - \frac{i}{2} 
  \left.
    \frac{d^2 \mathcal{Z}[\vec{f}, \vec{g}]}{%
      d f_i(\vec{q}) d g_j(-\vec{q})}
    \right|_{\vec{f},\vec{g} \to 0}.
\end{equation}

\begin{figure}[htpb]
  \centering
  \includegraphics[width=4in]{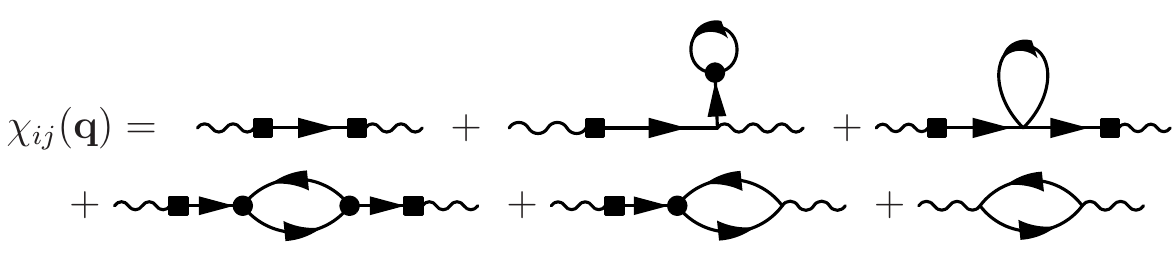}
  \caption{Feynman diagrams representing contributions to the response
    function at one-loop order.  Solid circles and squares involve
    factors of the quasicondensate density, and wavy lines represent
    coupling to currents.  The first five terms all contribute to the
    superfluid component, and the last gives the normal density.
    (Figure from Ref.~\cite{Keeling2011}.)}
  \label{keeling_fig:vertices}
\end{figure}

At this stage the calculation is exact; however, evaluating this
requires the ability to calculate the partition function exactly,
taking for the bare action an expression such as
Eq.~(\ref{keeling_eq:7}).  If a linearised approach is valid, one can
proceed by evaluating the path integral via a saddle point approach,
first minimising over $\bar \psi, \psi$, and then including quadratic
fluctuations about this saddle point.  For an equilibrium system to
respect sum rules, it is necessary that the saddle point should be
calculated \emph{in the presence of the fields $\vec f, \vec g$}.
Repeating this in the non-equilibrium case leads to a generating
function of the form $\mathcal{Z} \propto \exp\left(i S_0[\vec f, \vec g] -
\text{Tr} \ln\left( 1 + D A[\vec f, \vec g] \right)\right)$, with
contributions of the source terms to both the saddle point action and
the action for fluctuations.  Calculating the response function gives
a sequence of terms that can represented by the Feynman diagrams shown
in Fig.~\ref{keeling_fig:vertices}.  The first of these diagrams
corresponds to the contribution from the saddle point action $S_0$,
the others arise from terms such as $\text{Tr}[D (d^2 A/ d f_i d
g_j)]$ and $\text{Tr}[D (d A/d f_i) D (d A/d g_j)]$.

While the full expression coming from these diagrams~\cite{Keeling2011}
is rather involved, two simple conclusions can be drawn without
reference to these details.  The most important conclusion stemming
from the above formalism is that (as long as the linearised theory is
valid), a superfluid fraction will exist, due to the form of the first
five diagrams shown in Fig.~\ref{keeling_fig:vertices}.  Crucially, each of
these diagrams involves a term in which a single line carries the
entire incoming and outgoing momentum, and so they give expressions of
the form:
\begin{displaymath}
  \chi_{ij}^{SF}(\vec{q},\omega) \propto
  - |\psi_0|^2 q_i q_j  D^R(\vec{q},\omega),
\end{displaymath}
where $D^R(\vec{q},\omega)$ is the full retarded Green's function
(including the normal and anomalous self energies arising due to the
interaction terms in Eq.~(\ref{keeling_eq:7})).  This structure
reflects the fact that each diagram contains current vertices with one
line having zero momentum, while the other carries the full momentum
$\vec{q}$.  Hence each diagram includes a factor
$\gamma_i(\vec{q})\propto q_i$.  The retarded Green's function has
poles corresponding to the normal mode frequencies given in
Eq.~(\ref{keeling_eq:6}).  One may in fact show from the Keldysh
action that the Green's function has the form $D^R(\vec{q},\omega)
\propto C/[\omega(\omega+i\gamma_{\mathrm{net}}) - \xi_{\vec{q}}^2]$,
which has the crucial feature that $D^R(\vec{q}\to 0,\omega=0) \propto
1/q^2$.  This feature (along with the structure of the diagrams)
ensures a superfluid contribution to the response function.

The second important conclusion comes from the contribution
of the last diagram, giving the normal fraction.  This gives an
expression:
\begin{displaymath}
  m\chi_T
  = -\frac{i}{4}\int 
  \frac{d\omega}{2\pi}  
  \iint
  \frac{d^dk }{(2\pi)^d}  
  \epsilon_k 
  \text{Tr}\left[
    \sigma_z D^K_{\vec{k}}
    \sigma_z \left(D^R_{\vec{k}} + D^A_{\vec{k}}\right)
  \right],
\end{displaymath}
which can be evaluated, and shown not to vanish unless the loss terms
vanish.  This has a simple physical interpretation: the noise
associated with pumping and dissipation leads to the excitation of
quasiparticles, and thus the creation of a normal fraction in all
cases.

\subsection{Experimental probes of superfluidity}
\label{keeling_sec:exper-prob-superfl}

As emphasised in the preceding sections, the superfluid density
calculated above is a measure of the difference in how a system
responds to transverse (rotational) vs longitudinal forces. As such,
measurement of this superfluid response requires measuring the
response to a rotational perturbation of some kind.  For superfluid
Helium, the classic experiment is that of Andronikashvilli, using a
torsional oscillator formed of parallel discs, and studying the
changing inertia of the fluid as the temperature varies.

For polaritons, several issues arise: the particles are
quasiparticles, which do not strongly couple to a rotating inertial
frame, they live inside a semiconductor, and so applying a stirring
force is challenging, and their rotation does not provide any
measurable contribution to the mechanical angular momentum. What is
therefore required is a way to engineer an effective rotating frame
\emph{as seen by the polaritons}, and to measure the response of the
polaritons to this. 

Engineering a rotating frame for polaritons can be achieved in the
same spirit as proposed for cold atoms~\cite{Cooper2010,John2011},
namely manufacturing a real space Berry curvature arising from
spatially varying spin structure of polariton eigenstates.  The
essence of such a synthetic rotation is to use the two-component
spinor structure of the polariton to construct a spatially varying
ground state $|\chi(\vec r)\rangle$ such that the Berry connection
$\vec{A}(\vec{r}) = i \langle \chi | \nabla \chi \rangle$ takes the
form $\vec{A}(\vec{r}) = m \omega(r) \hat{\vec z} \times \vec{r}$,
corresponding to a synthetic rotating frame.  Such a configuration
arises from the ground state of the spin Hamiltonian $H = \lambda
[\ell_0^2 \sigma^z + r^2 (e^{2 i \theta} \sigma^- + \text{H.c.})]$,
where $r, \theta$ are in-plane  polar coordinates, and
$\sigma^i$ are Pauli matrices for the spinor basis.  The term $\lambda
\ell_0^2 \sigma^z$ corresponds to a Zeeman splitting induced by an
external field.  The other term requires either an induced strain, a
strong radial magnetic field (i.e. diverging in the plane of the
polariton system).  This leads to an angular velocity peaking at
$\omega \simeq 0.3 \hbar/m \ell_0^2$ around $r \sim \ell_0$.

Measuring the response to such a field is in principle possible in a
variety of ways, since an advantage of the polariton system is the
ability to directly probe polariton correlation functions in both real
and momentum space.  In particular, it is in principle possible to
measure correlations such as $\langle a^\dagger_{\vec k + \vec{q}}
a_\vec{k} \rangle$ as a function of $\vec k, \vec q$ by taking the
interference between two momentum-space images of the condensate,
displaced in real space, with a variable phase delay, i.e. calculating
$I(\phi_d) = \langle (a^\dagger_{\vec k} + e^{-i\phi_d}
a^\dagger_{\vec k + \vec{q}}) (a^{}_{\vec k} + e^{i\phi_d} a^{}_{\vec
  k + \vec{q}}) \rangle$, and mapping the fringe visibility as one
varies $\phi_d$.  Real space displaced fringe visibility maps are
routinely measured, see e.g.~\cite{Lagoudakis2008}; the equivalent
momentum space tomography would allow access to the current induced by
a given force, and thus reconstruction of the response function
$\chi_{ij}(\vec{q})$.

Existing experiments probing aspects of superfluidity in polaritons
are still  far from this limit.  Indeed what has been observed
thus far is suppression of drag, rather than the difference between
transverse and longitudinal response.  In addition, such experiments
to date have included measurements of the suppression of scattering
from defects, as a function of velocity and intensity of a coherently
driven condensate~\cite{Amo2009,amo09_b}.

\section{Vortices and metastable flow in a driven dissipative system}
\label{keeling_sec:vort-driv-diss}

As discussed in Section~\ref{keeling_sec:superfl-phase-coher}, the appearance
of quantised vortices demonstrates the existence of short range
coherence in a system, and is closely related to metastable persistent
flow in a non-simply-connected geometry.  We review here how these
features are modified in a dissipative condensate.  It is notable that
the structure of vortices changes in the presence of drive and
dissipation.  Because the dissipation depends on density, the
``continuity'' equation derived from Eq.~(\ref{keeling_eq:5}) takes the form:
\begin{equation}
  \label{keeling_eq:10}
  \partial_t \rho  + \nabla \cdot \vec j = (\gamma-\kappa - \Gamma \rho)\rho,
  \qquad
  \vec j \equiv - \frac{i}{2m} \left(
    \varphi^\ast \nabla \varphi
    -
    \varphi \nabla \varphi^\ast
  \right).
\end{equation}
This has the consequence that near the core of a vortex, where density
is depleted, there is net gain, and so there must be a
current  with non-vanishing divergence. Given that current
can be rewritten as $\vec j = \frac{\rho}{m} \nabla[
\text{arg}(\varphi)]$, this diverging current implies that vortices in
a driven dissipative system must have a spiral structure, with radial
as well as angular variation of phase. Such spiral vortices have been
discussed in the context of nonlinear optics~\cite{staliunas}.  The
existence of a spiral structure can modify the force between a pair of
vortices, and so may modify the nature of the vortex binding/unbinding
at the KT transition.  This provides a further complication --- in
addition to that provided by the strong nonlinearity of the KPZ
equation --- in understanding the KT transition in a driven
dissipative system.

There can also be cases in which combinations of spatial variation of
drive, dissipation, and potential trapping destabilise the vortex free
configuration, and instead stabilise configurations such as vortex
lattices~\cite{Keeling2008a,borgh10}. However, understanding whether
such configurations occur in practice requires analysis of the normal
state, going beyond the scope of the complex
GPE~\cite{Carusotto2013a}.

Because of the photon component of a polariton condensate, it is also
possible to directly imprint vortices on the condensate, by using a
coherent Gauss-Laguerre beam.  Calculations using stochastic classical
field methods have shown~\cite{Wouters2010a} how such pulses can
create metastable vortex states in both simply and
non-simply-connected geometries.  The additional noise associated with
pumping and decay means that the timescale for a vortex to move out of
such a condensate can be relatively short: rather than quantum
tunnelling, it can diffuse across a small annulus, driven by noise from
the pump and decay terms.

Vortices have been clearly observed in polariton
condensates~\cite{Lagoudakis2008}, however since most images of
polariton condensates require long integration times, vortices can
only be seen if they are either stationary (pinned on disorder), or if
they move along a repeatable path~\cite{Lagoudakis2011b} (so that
averages over many realisations recover the same trajectory).  As
such, indirect methods of imaging vortices may be necessary, such as
interference measurements with angular offsets~\cite{Wouters2010a}, or,
for vortex lattices, energy resolved and interference
images~\cite{borgh10}.

\section{Future directions, open questions}
\label{keeling_sec:future-direct-open}

The most profound open question is how the KPZ roughening interacts
with the fact that phase is a compact variable and can thus support
topological defects. As discussed previously two scenarios seem
possible. It may be that as the KPZ equation flows to strong
nonlinearity, the destruction of algebraic order also destroys any
resistance to vortex proliferation.  In this case, correlations are
always exponential, there is no ordered phase, no superfluidity, and
no phase transition.  In this scenario, all experiments on polariton
condensates would be finite size effects, although potentially
exceptionally large system sizes needed before the ``true'' behaviour
becomes visible.  The other possible scenario is that the strong
nonlinearity is compatible with vortex binding.  In this case, there
would be a phase transition between a high temperature phase showing
exponential decay of correlations, and a low temperature phase with
stretched exponential decay.  If this scenario holds then despite the
absence of algebraic order, there can be a non-zero superfluid
stiffness~\cite{Sieberer2015}. If such a phase transition exists, it
could potentially require a new universality class, distinct from the
equilibrium KT universality class.

While the above questions concern the fundamental physics of the
thermodynamic limit, a second set of questions concern the signatures
of this behaviour visible in finite experimental systems.
Physical~\cite{Roumpos2012,Nitsche2014} and
numerical~\cite{Dagvadorj2014} experiments have observed power law
correlations, but with surprising values of the power law exponent;
understanding the physical origin of this behaviour may help
understand the nature of the driven dissipative system.  Finally,
experiments directly probing the superfluid response function --- i.e.
measuring the superfluid fraction --- of a driven dissipative
condensate have yet to be realised.  Even in the context of cold atom
systems, direct measurements of the superfluid fraction remain
challenging~\cite{Cooper2010}.  Realising such measurements for
driven dissipative systems can provide confirmation and guidance to
the theoretical question of whether superfluidity exists in these
systems, and whether it is a finite-size or thermodynamic phenomenon.

\theendnotes

\end{document}